\documentclass[%
 aip,
 apl,%
 amsmath,amssymb,
 preprint,%
]{revtex4-1}

\usepackage{graphicx}
\usepackage{dcolumn}
\usepackage{bm}

\begin{document}


\title[Voltage controlled propagating spin waves on a perpendicularly magnetized nanowire]{Voltage controlled propagating spin waves on a perpendicularly magnetized nanowire}

\author{June-Seo Kim}
\affiliation{
Department of Applied Physics, Center for NanoMaterials, Eindhoven University of Technology, PO Box 513, 5600 MB Eindhoven, The Netherlands
}%
\author{Sanghoon Jung}
\affiliation{
Department of Physics, Sogang University, Seoul 121-742, Republic of Korea
}%
\author{Myung-Hwa Jung}
\affiliation{
Department of Physics, Sogang University, Seoul 121-742, Republic of Korea
}%
\author{Chun-Yeol You}
\email{cyyou@inha.ac.kr}
\affiliation{
Department of Physics, Inha University, Incheon 402-751, Republic of Korea
}%
\author{Henk J. M. Swagten}%
\affiliation{
Department of Applied Physics, Center for NanoMaterials, Eindhoven University of Technology, PO Box 513, 5600 MB Eindhoven, The Netherlands
}%
\author{Bert Koopmans}%
\affiliation{
Department of Applied Physics, Center for NanoMaterials, Eindhoven University of Technology, PO Box 513, 5600 MB Eindhoven, The Netherlands
}%

\date{\today}

\begin{abstract}
We numerically and analytically investigate the voltage controlled spin wave (SW) propagations in a nanowire with locally manipulated perpendicular magnetic anisotropy (PMA) by applying an electric field. It is shown that the velocity and wavelength of the propagating SWs are tailored by the modified PMA, which can be locally controlled by the external electric field. First, we observe a phase shift when the propagating SWs pass through the area of locally modified PMA. By introducing phase control of the SWs, we finally propose a three terminal SW device. Constructive/destructive interferences of two propagating SWs are controlled at the detecting area by the voltage controlled phase shift.
\end{abstract}

\maketitle

Spin wave (SW) always exists in magnetic systems at the finite temperature and has been investigated as fundamental research topics in the magnetic systems~\cite{Heinrich}. Of particular importance are the wave properties of the SWs such as reflection, transmission and interference~\cite{LeePRL2009,ChoiPRL2007,BalhornPRL2010}. SW based spintronic devices for ultrafast data processing, e.g., SW-based logic gates, have been numerically and experimentally realized by using the Mach-Zender type SW interferometer~\cite{SchneiderAPL2008,LeeJAP2008}. In Ref. 5, the functionality of SW logic device has been experimentally demonstrated with yttrium iron garnet (YIG) waveguides, but YIG is not compatible with standard silicon integrated circuit technology. Recently, the propagating SW induced magnetic domain wall (DW) motion has been theoretically and numerically demonstrated~\cite{HanAPL2009,SeoAPL2011,KimPRB2012,WangPRB2012}. Although it is a new concept to manipulate DWs, it has not been experimentally demonstrated and it is not compatible for spintronic devices due to extreme low DW propagation velocities~\cite{KimPRB2012,WangPRB2012}. On the other hand, recently, it has been reported that PMA can be controlled by an applied electric field due to the charge accumulations at the interface between metallic and oxide layers. It is able to open a new way to fabricate ferromagnetic field-effect devices~\cite{MaruyamaNN2009,SchellekensNC2012,BauerNN2013,FrankenAPL2013,HaAPL2010}. In this point of view, we are curious whether the wave properties of the propagating SWs can be also manipulated by controlling PMA due to charge accumulations or not. If a velocity of the propagating SW can be controlled by the electric field, it would be able to fabricate a radically different ultrafast SW logic device since the propagating velocity of the SW in a magnetic nanowire is higher than a few km/s~\cite{LeePRL2009,SekiguchiAPL2010}.

In this Letter, the wave properties of the propagating SW in a magnetic nanowire with locally controlled PMA values are analytically and numerically investigated by performing the micromagnetic simulations with the Object-Oriented MicroMagnetic Framework (OOMMF)~\cite{OOMMF}. According to our micromagnetic results, the propagating SWs velocities are changed by the PMA modulations. By changing PMA within a local area, we clearly observe a phase shift in the electric field region since the SW characteristics are changed as a function of the magnitude of PMAs. In other words, the SW dispersion relation is a function of the SW energy including the PMA. Finally, we propose a three terminal SW device that one can observe a destructive interference between two propagating SWs when the phase of one of two propagating SWs is changed by the electric field (or gate voltage). Since just few percentage changes of PMAs are enough to make significant phase shifts, this mechanism opens a new way to fabricate radically superior SW logic devices.

Figure~\ref{FigureDevice}(a) shows the schematic configuration ($xz$-plane) of the propagating SWs with a locally applied electric field. A blue wire indicate a nanowire with 2000-nm long, 100-nm wide, and 5-nm thick with PMA. The localized oscillating field ($B_{osc}$) is applied in the yellow area (width = 5 nm and length = 100 nm) and a green area indicates the localized electric field for controlling PMA. As expected, SWs are propagating along the $\pm x$ direction from the SW source (See Fig.~\ref{FigureDevice}(b)). The amplitude ($B_{osc}$) and frequency ($f_{H}$) of the sinusoidal oscillating field are fixed with $B_{osc}$ = 100 mT and $f_{H}$ = 13 GHz, respectively since the longest propagating SW is observed at $f_{H}$ = 13 GHz. The physical reason is explained later. We select a cell size of 2.5$\times$2.5$\times$2.5 nm$^3$ and standard material values with strong PMA are used in the simulations: the saturation magnetization $M_{s}$ = 8.6$\times$10$^5$ A/m, exchange stiffness $A_{ex}$ = 1.3$\times$10$^{-11}$ J/m, and the magnetic damping constant $\alpha$ = 0.01~\cite{WangPRB2012}. The dynamics of the propagating SW is governed by the Landau-Lifhshitz-Gilbert (LLG) equation:

\begin{equation} \label{LLG}
\frac{\partial \mathbf{M}}{\partial t} = -\gamma \mathbf{M} \times \mathbf{H}_{eff} + \frac{\alpha}{M_{s}}\mathbf{M}\times\frac{\partial \mathbf{M}}{\partial t},
\end{equation}

where $\mathbf{M}$ is the vector of local magnetization and $\gamma$ is the gyromagnetic ratio. $\mathbf{H}_{eff}$ is the effective field, which is composed of the exchange, anisotropy, magnetostatic, and external field. In order to generate monochromatic propagating SWs, we apply a strong harmonic sinusoidal external field $B_{osc}$ = $B_{0}\times \sin[2\pi f t]\mathbf{u_{y}}$ along the $y$ direction only at a localized area (5$\times$100$\times$5 nm$^3$) at $x$ = 500 nm from the left edge of the nanowire (See Fig.~\ref{FigureDevice}(a) and (b)). To prevent SW reflections from the edges, we include absorbing boundary conditions based on gradually increasing the magnetic damping~\cite{ConsoloIEEE2007}. Fig.~\ref{FigureDevice}(b) shows a snapshot ($xy$-plane) of the propagating SW at $t$ = 5 ns.

A propagating SW should be monotonously and exponentially attenuated as a function of position due to the damping effect. First, we calculate the amplitudes of the propagating SWs as a function of the position with several excitation frequencies and PMA values (See Fig.~\ref{SWpropagation}). At $x$ = 500 nm (SW source), we apply an oscillating field and then we monitor the $M_{x}/M_{s}$ value as a function of the simulation time ($t$ = 0 $\sim$ 10 ns). Since there is a magnetization fluctuation when a propagating SW arrives at a certain position, we only calculate the maximum and minimum values of the $M_{x}/M_{s}$ value during $t$ = 5 ns $\sim$ 10 ns to remove the fluctuation. Fig.~\ref{SWpropagation}(a) shows SW amplitudes ($\equiv ([M_{x}/M_{s}]_{max} - [M_{x}/M_{s}]_{min})/2$) as a function of the distance with various frequencies ($f_{H}$ = 5 GHz $\sim$ 20 GHz, and $\Delta f_{H}$ = 1 GHz). In Fig.~\ref{SWpropagation}(a), for the case of $f_{H}$ = 5 GHz (black squares), SWs are not able to propagate to the end of the nanowire since there is a forbidden gap of the SW dispersion relation~\cite{LeePRL2009,HanAPL2009}. On the other hand, for the cases of $f_{H}$ $>$ 9 GHz, there are long-lived propagating SWs up to $x$ = 1500 nm. Especially, we observe a monotonously and exponentially decreased propagating SW with $f_{H}$ = 13 GHz. Now we choose an excitation frequency $f_{H}$ = 13 GHz and then PMA value dependent SW amplitudes are calculated as a function of the position. For the case of the PMA $K_{u}$ $<$ 4.4$\times$10$^5$ J/m$^3$ and $K_{u}$ $>$ 6.2$\times$10$^5$ J/m$^3$, there is no propagating SW and we only observe a exponentially decreased propagating SWs with a certain range of the PMA values (5.51$\times$10$^5$ J/m$^3$ $<$ $K_{u}$ $<$ 5.58$\times$10$^5$ J/m$^3$, not visible). Fig.~\ref{SWpropagation}(b) shows the position dependent SW amplitudes with three different PMA values and the excitation frequency was fixed with $f_{H}$ = 13 GHz. Although, overall SW amplitude with $K_{u}$ = 5.6$\times$10$^5$ J/m$^3$ is larger than $K_{u}$ = 5.55$\times$10$^5$ J/m$^3$, but it is no longer monotonously decaying SW. Therefore, we find the optimized excitation frequency ($f_{H}$ = 13 GHz) and PMA value ($K_{u}$ = 5.6$\times$10$^5$ J/m$^3$). Hereafter, the PMA is fixed with $K_{u}$ = 5.6$\times$10$^5$ J/m$^3$.

In order to verify the correlation between propagating SWs and the magnitude of the PMA values, we calculate SW amplitudes as a function of position with locally changed PMA values (See Fig.~\ref{EField}(a)). The PMA values are locally changed between $x$ = 900 nm $\sim$ 1100 nm and we show snapshots of the propagating SWs at $t$ = 5 ns with three different PMA values (0.90 $K_{u}$, 0.95 $K_{u}$, 1.00 $K_{u}$, 1.05 $K_{u}$, and 1.10 $K_{u}$, respectively) changed by the locally applied electric fields. As shown in Fig.~\ref{EField}(a), the wavelengths of the propagating SWs are changed at the area of the electric field. For the case of lower (higher) PMA value with 0.90 $K_{u}$ and 0.95 $K_{u}$ (1.05 $K_{u}$ and 1.10 $K_{u}$), the wavelength of the SWs obviously decreases (increases). Fig.~\ref{EField}(b) shows the calculated $M_{x}/M_{s}$, which is directly proportional to the SW amplitude, as a function of the position ($x$ = 800 nm $\sim$ 1400 nm). We also observe that the wavelengths of the SWs are clearly tuned while the propagating SW travels through the electric field area and the wavelength of the propagating SW comes back to the initial excited SW with different phase shifts (PSs). The calculated PSs for both PMA values (0.95 $K_{u}$ and 1.05 $K_{u}$) are 1.17$\pi$ and 0.98$\pi$, respectively. Such changes of the wavelength and the velocity have same physics with the general propagating wave in two different media~\cite{XiJAP2008}.

This change of the propagating SW characteristics by the PMA value can be understood from the dispersion relation for the PMA system with a confined geometry~\cite{Heinrich}:

\begin{eqnarray} \label{DE}
\omega_{res} &=& \gamma^{2}[H_{0}+\frac{2K_{u}}{\mu_{0}M_{s}}+(N_{y}-N_{z})M_{s}+\frac{2A_{ex}}{\mu_{0}M_{s}}k^2] \nonumber \\
&\times& [H_{0}+\frac{2K_{u}}{\mu_{0}M_{s}}+(N_{x}-N_{z})M_{s}+\frac{2A_{ex}}{\mu_{0}M_{s}}k^2]
\end{eqnarray}

where, $\omega_{res}$ is the resonance frequency, $H_{0}$ is the in-plane applied field, $K_{u}$ is the PMA energy, $\mu_{0}$ is the permeability of the vacuum, and $N_{x}$, $N_{y}$, and $N_{z}$ are demagnetization factors for $x$, $y$, and $z$ axes, respectively. Now, we ignore the dipolar SW interaction (Damon-Eshba mode)~\cite{DamonJCPS1961} and assume that PMA values are linearly changed with the electric field $E$, $K_{u}$ = $K_{u}(E)$ + $\Delta K_{u}E$~\cite{SchellekensNC2012}. By using a first order approximation, we can obtain a simple expression between PMA value and the wavevector of the propagating SWs:

\begin{eqnarray} \label{Dispersion}
\Delta k &=& -\frac{\Delta K_{u}}{2A_{ex}k(0)}, \nonumber \\
v_{SW}&=&\frac{\omega_{res}}{k(0)+\Delta k E}=\frac{\omega_{res}}{k(0)-\frac{\Delta K_{u}}{2A_{ex}k(0)}E} \nonumber \\
 &\approx& v_{SW}\left [ 1+\frac{\Delta K_{u}}{2A_{ex}k^2_{0}}E \right]
\end{eqnarray}

It should be noted that the SW frequency does not change with the variation of the PMA in Eq.~\ref{Dispersion}. This fact indicates that the wavevector of the propagating SW decreases with a higher PMA due to applying electric field as shown in Fig.~\ref{EField}(b). And we find that the SW velocity increases with the increasing PMA value (or electric field with an assumption of positive $\Delta k_{u}$) in Eq.~\ref{Dispersion}. Therefore, we convince that the velocity of the propagating SW increases with increasing wavelength. Fig.~\ref{DispersionGraph}(a) shows the $M_{x}/M_{s}$ value with two different times ($\Delta t$ = 0.02 ns) and we now determine the velocity of the SW ($v_{SW}$ = $\Delta x / \Delta t$). In Fig.~\ref{DispersionGraph}(b), we plot the velocities of the propagating SWs as a function of the PMA. Black circles and red circles indicate the SW velocities from the micromagnetic simulations and the dispersion relation in Eq.~\ref{DE}, respectively. Blue dashed line shows the SW velocities from the first order approximation in Eq.~\ref{Dispersion}. The simulated SW velocities are in good agreement with the analytical calculations. The small discrepancies are due to the uncertainties of the demagnetization factors for the nanowire.

Finally, we can propose a three terminal SW device with PMA manipulation by a localized applying electric field. Fig.~\ref{Threeterminal}(a) shows the schematic configuration of the three terminal SW device. Two parallel ferromagnetic nanowires (100 nm width and 100 nm spacing) are shown here and these nanowires are merged after $x$ = 800 nm. The total length of the nanowire is 2000 nm. The localized oscillating field ($B_{osc}$ = 100 mT) is applied at $x$ = 200 nm (SW source), with $f_{H}$ = 13 GHz. The yellow area (400 nm $<$ $x$ $<$ 700 nm) indicates the localized electric field and the SW is measured at $x$ = 1800 nm. Now, we change the strength of the PMA from 0.90 $K_{u}$ to 1.10 $K_{u}$ on the electric field area. In Fig.~\ref{Threeterminal}(b), two propagating SWs are constructively (in-phase) with PMA 1.00 $K_{u}$ interfered due to the symmetry of the system. On the other hand, for the cases of PMA with 0.96 $K_{u}$ and 1.03 $K_{u}$, two SWs are destructively (out-of-phase) interfered since the phase shifts between two propagating SWs are almost $\pi$ (See Fig.~\ref{Threeterminal} (b $\sim$ d)). The underlying physics are exactly same with the optical interferences. The key idea of our present work is the phase shift due to the locally manipulated SW wavelengths, or velocity, which is controllable by the external electric field. Fig.~\ref{Threeterminal}(e) shows the $M_{x}/M_{s}$ values as a function of the simulation time ($t$ = 5.0 ns $\sim$ 5.5 ns) at $x$ = 1800 nm for two different PMA values. Compare with the case without an electric field (1.00 $K_{u}$), the SW amplitudes at $x$ = 1800 nm with 0.96 $K_{u}$ is approximately 5 times smaller. Now, we emphasize that the amplitude of the propagating SWs at the end of the nanowire is almost vanished even with only a few percent change of PMA strength in the upper nanowire. Fig.~\ref{Threeterminal}(f) shows the interfered SW amplitudes at $x$ = 1800 nm as a function of the locally modified PMA values. As expected, the SW amplitudes at the detecting area are drastically changed as the PMA values vary from 0.91 $K_{u}$ to 1.09 $K_{u}$. Moreover, an oscillation of the SW amplitude is observed since the wavelength of the propagating SW is linearly changed. It should be noted that this approach works for ultrafast (over 10 GHz) new logic devices based on propagating SWs.

In conclusion, we analytically and numerically investigate the SW logic devices based on the electric field control PMA. We observe that the wavelength and velocity of the propagating SW can be controlled by the locally manipulating PMA, and consequently a controllable phase shift (and interference) is introduced with the electric field in a specific area. The analytical expression reveals that the wavevector of the propagating SWs are changed with the strength of the PMA. By introducing this electric field controlled PMA mechanism, we finally propose a three terminal SW device. In the proposed device, the amplitudes of the propagating SWs are well controlled by the constructive/destructive interferences which are manageable by the external local electric field.

This work is supported by the research programme of the Foundation for Fundamental Research on Matter (FOM), which is part of the Netherlands Organisation for Scientific Research (NWO), and NRF funds (Grant Nos. 2013R1A1A2011936 and 2012M2A2A6004261) and by the IT R\&D program of MKE/KEIT [10043398] of Korea.

\newpage
\label{FigureDevice}  Figure 1. (a) Schematic configuration ($xz$-plane) of the voltage controlled SW propagation. The blue rectangle indicates a nanowire (length = 2000 nm, width = 100 nm, and thickness = 5 nm) with PMA of $K_{u}$ = 5.55$\times$10$^5$ J/m$^3$. An oscillating magnetic field ($B_{osc}$ = 100 mT) along the $y$-direction is applied in the green region (width = 5 nm and thickness = 5 nm). The red rectangle indicates the top electrode for locally changed PMA. (b) The snapshot ($xy$-plane) of the propagating SW with $f_{H}$ = 13 GHz and $B_{osc}$ = 100 mT at $t$ = 5 ns. \\

\label{SWpropagation}  Figure 2. The SW amplitudes as a function of the distance with various SW excitation frequency ($f_{H}$ = 5 GHz and 12 $\sim$ 14 GHz). The SW excitation field is fixed at $B_{osc}$ = 100 mT. (b) The SW amplitudes as a function of the distance with three different PMA values ($K_{u}$ = 5.50$\times$10$^5$, 5.55$\times$10$^5$, and 5.60$\times$10$^5$ J/m$^3$). The SW excitation frequency is fixed at $f_{H}$ = 13 GHz. \\

\label{EField}  Figure 3. (a) Snapshots of propagating SWs with different PMAs (0.90 $K_{u}$, 0.95 $K_{u}$, 1.00 $K_{u}$, 1.05 $K_{u}$, and 1.10 $K_{u}$) due to the localized electric field (900 nm $<$ $x$ $<$ 1100 nm). (b) The calculated $M_{x}/M_{s}$ as a function of the distance ($x$ = 800 nm $\sim$ 1400 nm) at $t$ = 5 ns. PS indicates the phase shift between propagating SWs with manipulated PMAs (0.95 $K_{u}$ and 1.05 $K_{u}$) and propagating SW without an electric field effect (1.00 $K_{u}$). \\

\label{DispersionGraph}  Figure 4. (a) The calculated $M_{x}/M_{s}$ as a function of the distance ($x$ = 750 nm $\sim$ 1250 nm) at $t$ = 5.001 ns and $t$ = 5.021 ns with a PMA of $K_{u}$. The SW excitation frequency is $f_{H}$ = 13 GHz and amplitude is $B_{osc}$ = 100 mT. (b) The propagating SW velocities as a function of the PMA values (0.95 $K_{u}$ $\sim$ 1.05 $K_{u}$). Black circles and red circles indicate the propagating SWs velocities from micromagnetic simulations and dispersion relation in Eq.~\ref{DE}, respectively. Blue dashed line shows the SW velocities from the first order approximation (See Eq.~\ref{Dispersion}).  \\

\label{Threeterminal} Figure 5. (a) Schematic configuration of a three terminal SW logic device. A 2000 nm long and 100 nm wide three terminal nanowire is examined. Two branches are merged after $x$ = 800 nm. The yellow area ($\Delta x$ = 300 nm) indicates the area of the applied electric field. SWs are generated at $x$ = 200 nm and detected at $x$ = 1800 nm. (b)-(d) Snapshots of the propagating SWs with different PMA values (b) 1.00 $K_{u}$, (c) 0.96 $K_{u}$, and (d) 1.03 $K_{u}$ at $t$ = 5 ns. The amplitude ($B_{osc}$) and frequency ($f_{H}$) of the propagating SWs are fixed with 100 mT and 13 GHz, respectively. Due to a symmetric excitation and propagation, two propagating SWs are merged in phase for the case of 1.00 $K_{u}$. For the cases of 0.96 $K_{u}$ and 1.03 $K_{u}$, the SW amplitudes are destructively interfering each other since they are in out-of-phase. (e) Calculated $M_{x}/M_{s}$ as a function of time ($t$ = 5.0 ns $\sim$ 5.5 ns) with various PMA values at $x$ = 1800 nm. Blue line and red line are results with the strengths of the PMA 0.96 $K_{u}$ (out-of-phase) and 1.00 $K_{u}$ (in-phase), respectively. (f) SW amplitudes at $x$ = 1800 nm. The locally changed PMA values vary from 0.90 $K_{u}$  to 1.10 $K_{u}$.

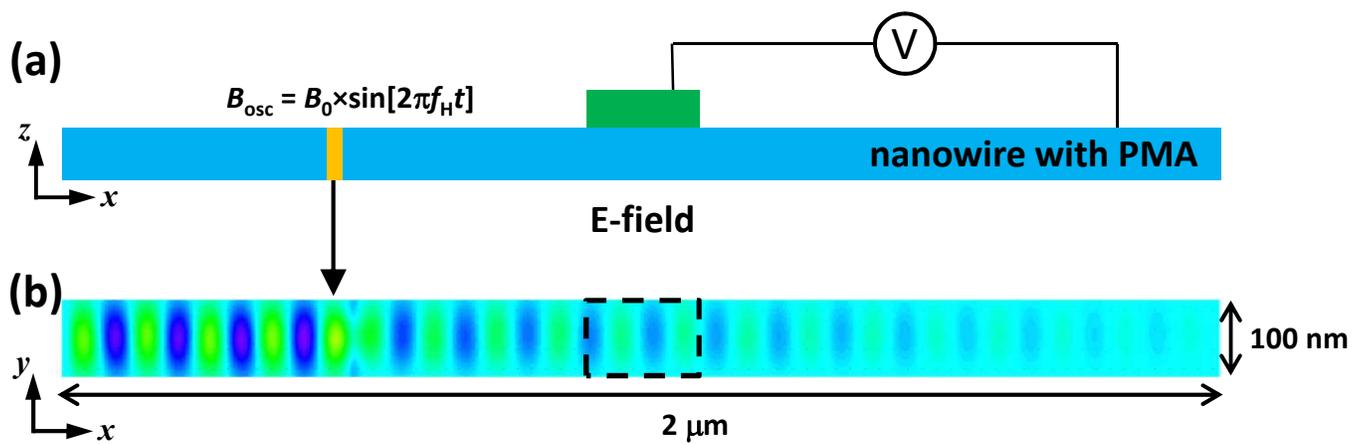
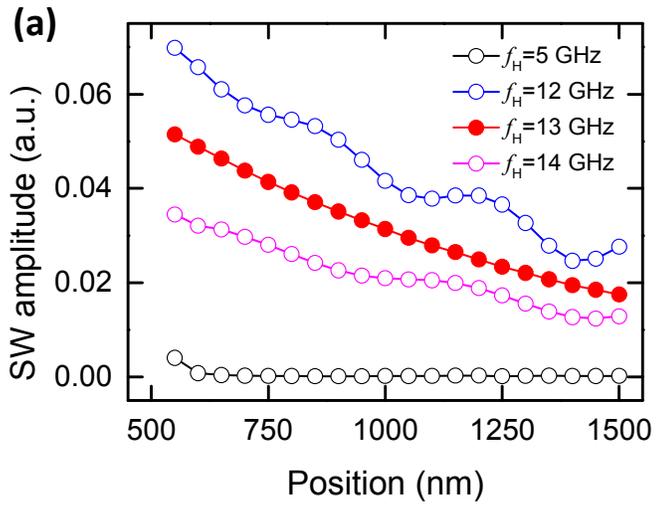
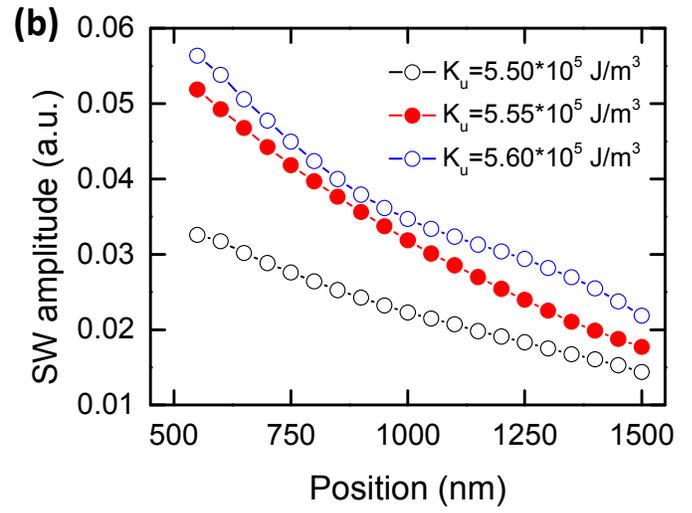
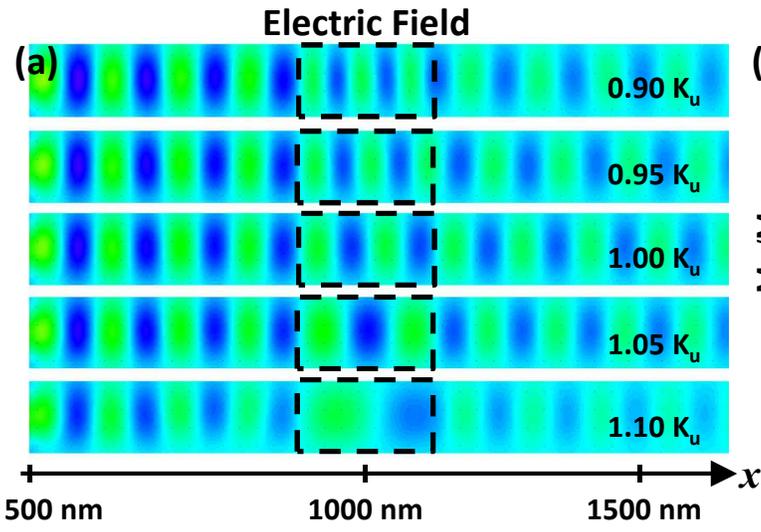
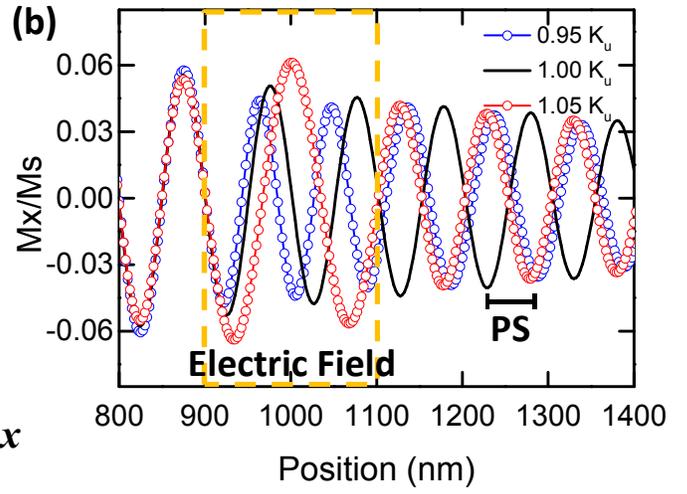
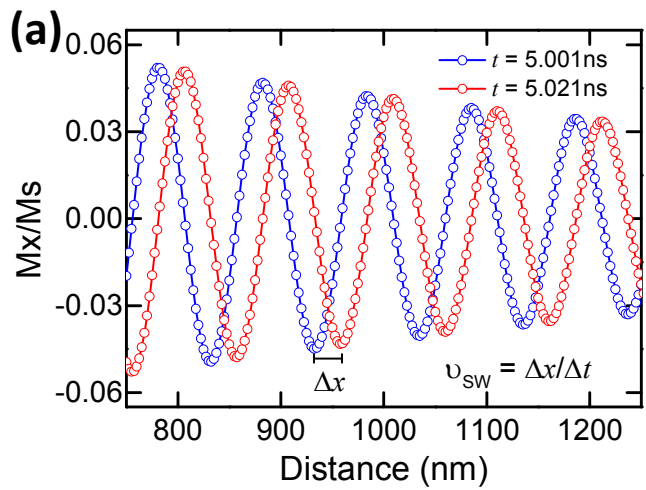
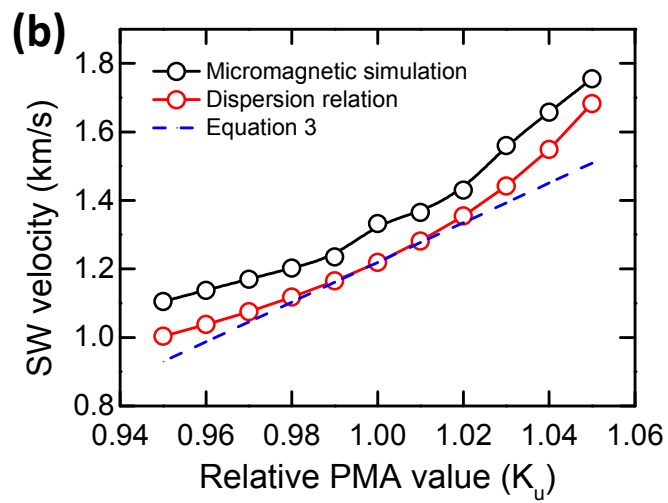
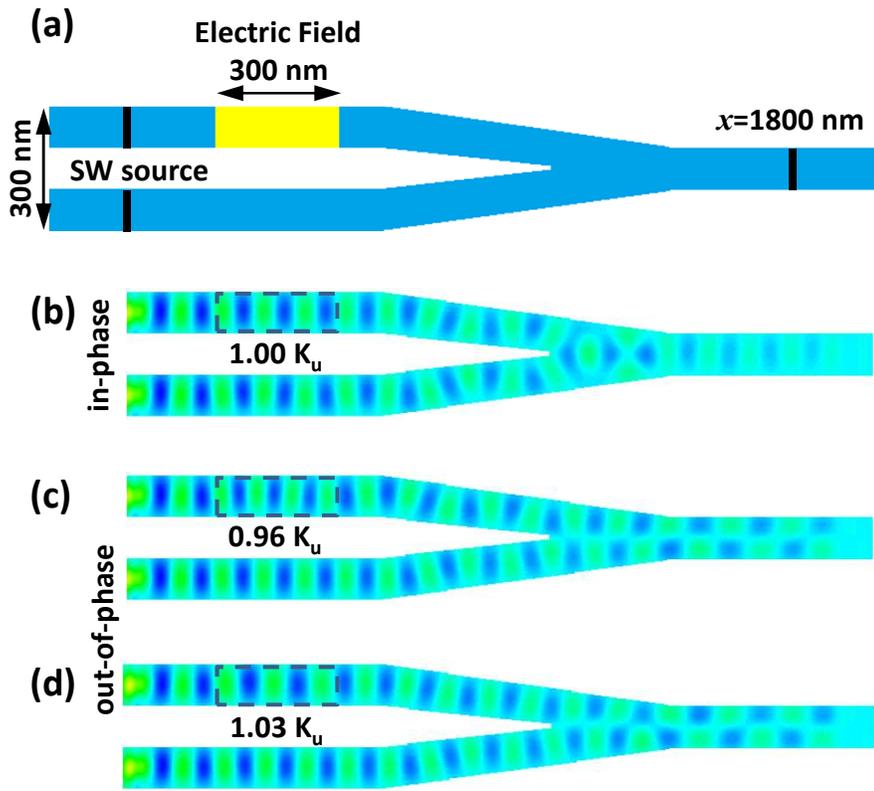
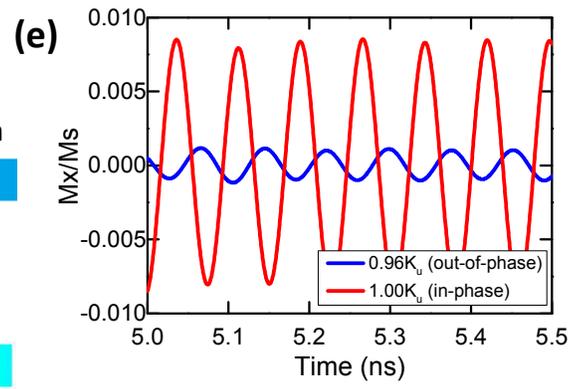
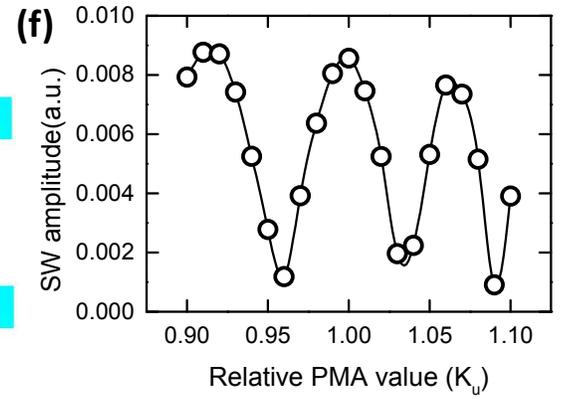
\end{document}